\documentclass[10pt,a4paper]{article}
\usepackage[top=100pt,bottom=100pt,left=60pt,right=60pt]{geometry} 
\usepackage[sort,compress]{cite} 

\usepackage{graphicx} 
\usepackage{amsmath} 
\usepackage{amssymb} 
\usepackage{cases} 
\usepackage{setspace} 
\usepackage{color} 
\usepackage{textcomp} 
\usepackage{caption} 
\usepackage{url}

\usepackage{array} 
\newcolumntype{x}[1]{>{\centering\arraybackslash\hspace{0pt}}m{#1}}

\title{\textbf{\large{Spatiotemporal Oscillations Driven by Spatial Heterogeneity and Growth}}}

\author{\normalsize{L.S. Mosby$^{1,2,3,*}$, M. P. Dalwadi$^{4,\dagger}$, Z. Hadjivasiliou$^{1,2,3,\dagger}$} \\
	\small\textit{$^1$The Francis Crick Institute: Mathematical and Physical Biology Laboratory,} \\
	\small\textit{1 Midland Road, London, NW1 1AT, UK} \\
	\small\textit{$^2$University College London: Department of Physics and Astronomy,} \\
	\small\textit{Gower Street, London, WC1E 6BT, UK} \\
	\small\textit{$^3$London Centre for Nanotechnology, 19 Gordon Street, London, WC1H 0AH, UK} \\
	\small\textit{$^4$Mathematical Institute, University of Oxford, Woodstock Road, Oxford, OX2 6GG, UK}} 

\begin{document}

    \maketitle
    \def\thefootnote{*}\footnotetext{Corresponding author, email address: l.mosby@ucl.ac.uk}
    \def\thefootnote{$\dagger$}\footnotetext{These authors contributed equally to this work}

    \begin{spacing}{1.8}

    \begin{abstract}

    Temporal oscillations arise naturally in biological systems with feedback loops or explicit delays. When incorporating spatial effects, the travel times of propagating signals can also generate intrinsic delays, and this behaviour is complicated further in the presence of spatial heterogeneity. In this work, we investigate the mechanistic origin and dynamics of spatiotemporal oscillations driven by intrinsic diffusive delays in a representative activator-inhibitor system with spatial heterogeneity using a combination of mathematical modelling, simulations and numerical methods. In contrast to delay-differential equation models where delays are explicitly prescribed, in the partial differential equation system we consider here delays are an intrinsic consequence of signal travel time through space. We demonstrate that undamped oscillations can arise as a result of spatially heterogeneous feedback when the system size increases through a critical length. At this critical length the time for systems to reach steady-state diverges, and the systems first exhibit an eigenvalue with a positive real part and non-zero imaginary part, corresponding to a Hopf bifurcation. We infer that this critical length could define a theoretical upper limit for the size of systems where oscillations hinder function. The analysis pipeline derived in this work offers a novel route for understanding the onset of spatiotemporal oscillations and Hopf bifurcations in a diverse array of biological systems where interactions are mediated by diffusive species.

    \end{abstract}

    \pagebreak

    \section{Introduction}

    Temporal oscillations are a natural consequence of interactions mediated by feedback loops, and are present in many biophysical systems \cite{fussmann2000crossing, martinez2018bistable, di2018landau, kuniya2025hopf, alonso2025amplification,ridgway2023motility}. These oscillations can also be driven by delays, which offset and prevent the timely response of feedback \cite{macnamara2016diffusion, martinez2018bistable, biga2021dynamic, glass2021nonlinear}. A Hopf bifurcation characterises the emergence of limit cycles, corresponding to states of indefinite oscillations, that arise when a parameter is varied through a critical value in certain systems \cite{fussmann2000crossing, martinez2018bistable, di2018landau, biga2021dynamic, glass2021nonlinear,ridgway2023motility}. Qualitatively, this can occur when timescales associated with feedback become slower than those associated with system dynamics, resulting in the over-correction of perturbations. A Hopf bifurcation can cause loss of function in biological systems, where timely responses to stimuli are often necessary. However, there are cases where undamped oscillations are useful; for example, persistent HES5 oscillations are necessary for neuronal differentiation, and arise in models with self-repressive feedback and transcriptional delays \cite{macnamara2016diffusion, biga2021dynamic}.
    
    The addition of a spatial component to the already complex dynamics of temporal oscillations and limit cycles make spatiotemporal oscillations significantly more challenging to analyse. For example, introducing spatial heterogeneity into the Gierer-Meinhardt equation results in travelling spike solutions that are not amenable to analysis using shadow limit asymptotic methods \cite{krause2018heterogeneity}. In biological systems, signals can be propagated by cell-cell interactions \cite{martinez2018bistable} or diffusing molecules \cite{ben2010scaling, macnamara2016diffusion, almuedo2018scale, mosby2026position}, and signal travel time can lead to intrinsic delays. For example, stress oscillations in growing bacterial biofilms arise due to signalling delays between cell-cell contact interactions from the periphery and the interior \cite{martinez2018bistable}. This behaviour can be modelled using delay-differential equations (DDEs) \cite{martinez2018bistable, glass2021nonlinear}. However, since delays are imposed, and not emergent, properties of DDE models, a DDE approach obscures the mechanistic origin of spatiotemporal oscillations by collapsing intermediate steps into a single constitutive (lumped) parameter. In contrast, HES1 protein oscillations necessary for somitogenesis can be reproduced qualitatively in simulations with intrinsic delays that arise due to the time it takes mRNA and proteins to diffuse between point-like, spatially-separated interaction sites \cite{macnamara2016diffusion}. However, biological systems often exhibit spatially-extended (non-point-like) regions where interactions occur between diffusing species. In this work, we investigate how intrinsic delays lead to the emergence of spatiotemporal oscillations in spatially heterogeneous systems where interactions can occur at any position.
    
    We will tackle this problem in the context of activator-inhibitor feedback, a form of negative feedback ubiquitous, for example, across the field of developmental biology \cite{bastida2009bmp, raspopovic2014digit, almuedo2018scale}, where signals are transmitted by diffusing molecules. Notably, this type of feedback has been observed in the expansion-repression (ER) model, which has been attributed to tissue pattern scaling during development \cite{ben2010scaling, mosby2026position}. It was previously hypothesised that spatiotemporal oscillations arise in the ER model due to molecular diffusion introducing a delay element into the established integral-feedback control mechanism \cite{ben2010scaling}. However, this proposed mechanism does not predict the onset of undamped oscillations observed in recent work \cite{mosby2026position}. This raises important questions about the mechanistic origin of undamped spatiotemporal oscillations in systems with activator-inhibitor feedback, and whether these oscillations could help or hinder patterning during development.

    In this work, we first demonstrate analytically that undamped spatiotemporal oscillations in our activator-inhibitor system must be driven by spatial heterogeneity. We quantify using simulations that the time to reach steady-state diverges when system size approaches a critical length, indicating that the transition from damped to undamped oscillations represents a Hopf bifurcation. Additionally, we find strong correlation between the oscillation period and the diffusive timescales of the activator and inhibitor molecules travelling between their respective source regions, linking the bifurcation to intrinsic diffusion-driven delays coupled with spatial heterogeneity. We demonstrate numerically that our activator-inhibitor feedback generates an eigenvalue with a positive real part when the system size exceeds the critical length obtained from our simulations, and phenotypes including both super- and sub-critical Hopf bifurcations. Finally, we use a simplified DDE model to analytically derive a delay threshold that induces a Hopf bifurcation, mirroring our findings for systems with strongly spatially heterogeneous activator-inhibitor feedback. We conclude by linking our results to previously published work highlighting the benefits of operating near criticality for biological systems \cite{di2018landau, zhang2022fgf8, alonso2025amplification}.

    \section{Results}
    
    \subsection{Mathematical model} \label{sec:ModelSystem}

    \begin{figure}[h!]
        \centering
            \includegraphics[width = 0.75\linewidth]{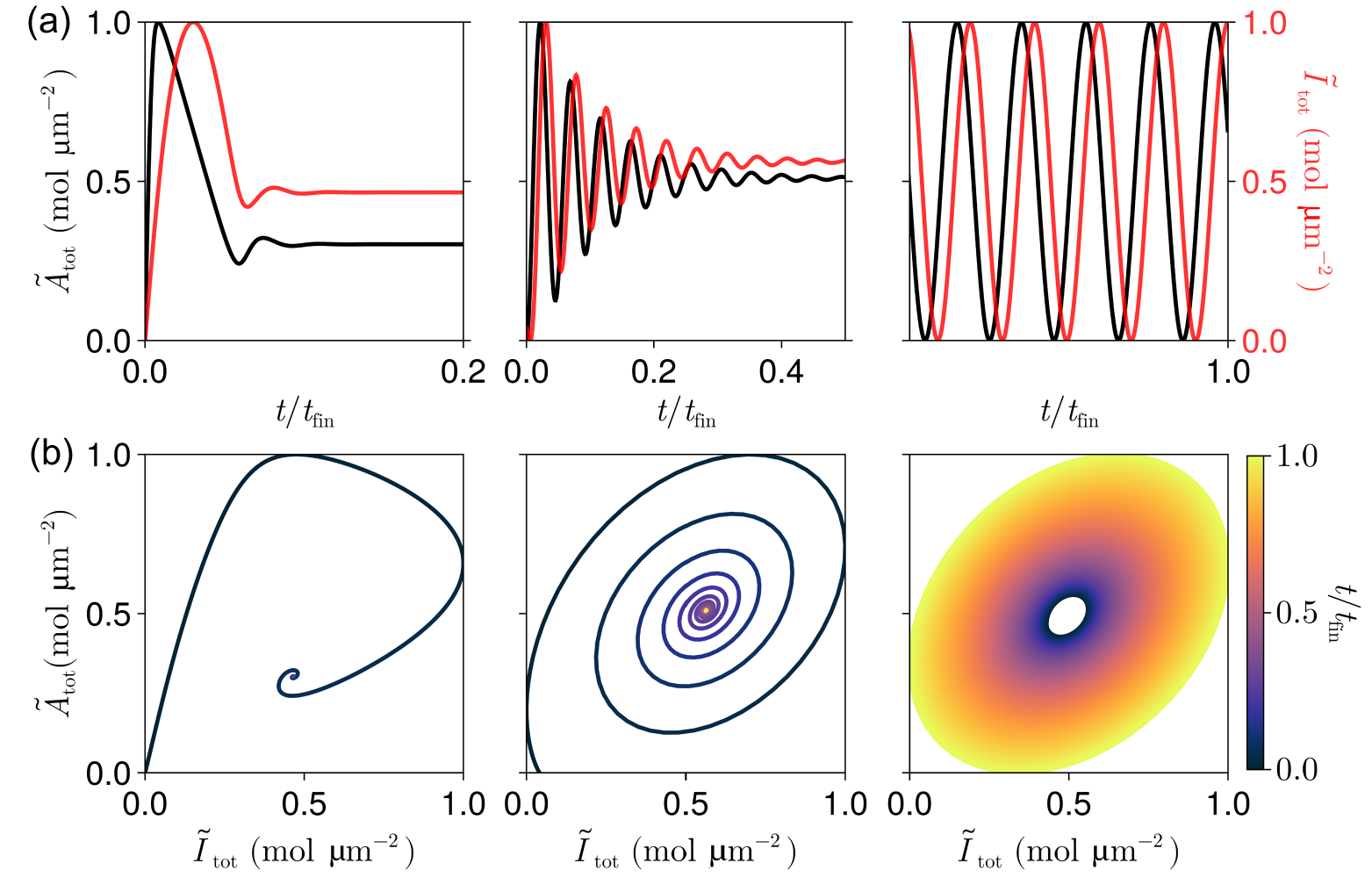}
            \caption[Spatiotemporal oscillations in systems with activator-inhibitor feedback]{\textbf{Spatiotemporal oscillations in systems with activator-inhibitor feedback.} (a) The total activator ($A_\text{tot} = \int_x A \, dx$, $\tilde{A}_\text{tot} = (A_\text{tot} - \min(A_\text{tot})) / (\max(A_\text{tot}) - \min(A_\text{tot}))$ where max and min correspond to data within the plotted time window) and inhibitor (calculated using the same method as for $A_\text{tot}$) concentrations for representative systems obtained by simulating eqs.~(\ref{eq:FP_a},\ref{eq:FP_b}), plotted as a function of fractional simulation time $t/t_\text{fin}$, where $t_\text{fin}$ is the termination time of the simulation and varies between simulations. (b) The corresponding data plotted in total concentration-space, where the colour indicates the fractional simulation time. System parameters are presented in tables in Appendix \ref{sec:SIMethods}.}
            \label{fig:Oscillations}
    \end{figure}

    In this work, we focus on the expansion-repression (ER) feedback network \cite{ben2010scaling, mosby2026position}, where feedback is mediated through both degradation and production without any explicit delays. For this reason, we operate on the spatial and temporal scales of microns and seconds, respectively (parameter values are presented in tables in Appendix \ref{sec:SIMethods}). However, we note that the following derivations and analysis are applicable to a diverse array of biological systems where feedback is mediated by diffusive molecules. In order to maintain generality, we define the ER model in terms of activator ($A$) and inhibitor ($I$) species, which are coupled through the equations,
    \begin{align}
        \partial_t A &= D_A \partial_x^2 A - \left( \frac{k_A \, \alpha}{\alpha + A} \right) A + \nu_A \left( \frac{\beta^h}{\beta^h + I^h} \right), \label{eq:FP_a} \\
        \partial_t I &= D_I \partial_x^2 I - \left( \frac{k_I \, \gamma}{\gamma + A} \right) I + \nu_I \, \theta(w_I - x), \label{eq:FP_b}
    \end{align}
    \noindent where $D_{A,I}$ are the activator and inhibitor diffusivities, respectively, $k_{A,I}$ are the upper limits of the corresponding degradation rates, $\nu_{A,I}$ are the corresponding production rates, $\theta(\cdot)$ is the Heaviside step function, $w_I$ is the width of the inhibitor source region, and $\alpha$, $\beta$, $\gamma$ and $h$ are constants. We assume that the inhibitor source width $w_I$ scales linearly with the system size $L$ ($w_I \sim L$), and impose zero diffusive flux boundary conditions,
    \begin{equation}
        \partial_x A|_{x=0,L} = 0, \qquad \partial_x I|_{x=0,L} = 0.
        \label{eq:BoundaryConditions}
    \end{equation}
    \noindent Finally, we assume that $A$ and $I$ relax to steady-state quickly compared to characteristic timescales associated with changes in system size, such that the system is able to reach a stable fixed point or limit cycle at each value of $L$. This allows us to probe system stability for different system sizes.    

    Simulations of eqs.~(\ref{eq:FP_a},\ref{eq:FP_b}) reveal that spatiotemporal oscillations emerge under certain conditions (Fig.~(\ref{fig:Oscillations}); see Appendix \ref{sec:SIMethods} for methods). For some sets of feedback and system parameters these oscillations are damped, resulting in the system tending towards a stable fixed point in total concentration-space (left and middle columns of Fig.~(\ref{fig:Oscillations})) \cite{ben2010scaling, mosby2026position}. For other sets of feedback and system parameters undamped oscillations are observed, corresponding to limit cycles in total concentration-space (right column of Fig.~(\ref{fig:Oscillations})) \cite{mosby2026position}. Our goal is to generate a mechanistic understanding of how these oscillations arise and transition from being damped to undamped. In order to achieve this, we will show that the transition arises due to a Hopf bifurcation and subsequently analyse the onset of this bifurcation.

    \subsection{Undamped oscillations require spatial heterogeneity} \label{sec:Turing}

    We first establish that spatial heterogeneity is important for the onset of undamped oscillations in the activator-inhibitor system defined by eqs.~(\ref{eq:FP_a},\ref{eq:FP_b}). We do this by testing whether either of the two homogeneous steady-states, corresponding to the two different values of the Heaviside function in eq.~(\ref{eq:FP_b}), exhibit Turing instabilities \cite{turing1952chemical}. For this derivation, we consider small perturbations of the form $\underline{\delta u} = (\delta A, \delta I)$ around either fixed point $\underline{\bar{u}} = (\bar{A}, \bar{I})$. The dynamics of the perturbations $\underline{\delta u}$ can be approximated via the linearisation $\partial_t \, \underline{\delta u} \simeq \underline{\underline{J}} \, \underline{\delta u}$, where $\underline{\underline{J}}$ is the Jacobian of eqs.~(\ref{eq:FP_a},\ref{eq:FP_b}). The Jacobian of eqs.~(\ref{eq:FP_a},\ref{eq:FP_b}) is,
    
    \begin{equation}
    \underline{\underline{J}} = \begin{bmatrix}
            -\left( \frac{k_A \, \alpha^2}{\left(\alpha+\bar{A}\right)^2} \right) + D_A \, \partial_x^2 & - \left( \frac{\nu_A \, h \, \beta^h \, \bar{I}^{h-1}}{\left(\beta^h + \bar{I}^h\right)^2} \right) \\
            \frac{k_I \, \gamma \, \bar{I}}{\left(\gamma+\bar{A}\right)^2} & -\left( \frac{k_I \, \gamma}{\gamma+\bar{A}} \right) + D_I \, \partial_x^2
        \end{bmatrix}.
        \label{eq:Jacobian}
    \end{equation}
    
    In the absence of diffusion ($D_{A,I} = 0$), $\text{Tr}(\underline{\underline{J}}) < 0$ and $\text{Det}(\underline{\underline{J}}) > 0$ for the Jacobian defined in eq.~(\ref{eq:Jacobian}). In this case, the corresponding eigenvalues always exhibit a negative real part and the system is stable (see Appendix \ref{sec:SI1}). More specifically, damped oscillations arise when $\text{Tr}(\underline{\underline{J}})^2 - 4 \, \text{Det}(\underline{\underline{J}}) < 0$, which occurs when $J_{1,2} \, J_{2,1} < -((J_{1,1}-J_{2,2})/2)^2$. This mirrors the dynamics of the total activator level $A_\text{tot}$ observed in the left and middle columns of Fig.~(\ref{fig:Oscillations}).
    
    In the presence of diffusion ($D_{A,I} > 0$), we look for wave-like solutions of the form $\underline{\delta u} = \sum_i \underline{v}_i \exp(\lambda_i \, t + i \, q_i \, x)$. In this case, calculation of the eigenvalues of eq.~(\ref{eq:Jacobian}) demonstrates that the system is also stable in the presence of diffusion ($\text{Re}(\lambda_i) < 0$ $\forall i$), since the system cannot support real wavenumbers $q_i$ for a value of $\text{Det}(\underline{\underline{J}}) < 0$, a condition necessary for instability (see Appendix \ref{sec:SI1}).
    
    Our results suggest that the observed instability of the activator-inhibitor system defined by eqs.~(\ref{eq:FP_a},\ref{eq:FP_b}), which drives the system from damped to undamped oscillations (Fig.~(\ref{fig:Oscillations})), cannot be explained by considering only the instabilities associated with spatially homogeneous base states. In the following sections we probe the effects of spatial heterogeneity on system stability using a combination of simulations and numerical methods.

    \subsection{Spatiotemporal oscillations are driven by intrinsic diffusive delays} \label{sec:SimResults}

    Since we demonstrated in the previous section that spatially homogeneous base states are stable to perturbations, we hypothesised that the transition from damped to undamped oscillations could be a property of spatially heterogeneous systems due to the travel time associated with perturbations in activator or inhibitor levels diffusing between different spatial regions. Whereas the explicit treatment of delays has been shown to induce temporal oscillations \cite{macnamara2016diffusion, martinez2018bistable, biga2021dynamic, glass2021nonlinear}, the travel time here would instead introduce a source of intrinsic delays. For this reason, we used the system size $L$ as a control parameter to understand the dynamics of spatiotemporal oscillations, since increasing the distance between different spatial regions should also increase the magnitude of the intrinsic signalling delays.

    We simulated eqs.~(\ref{eq:FP_a},\ref{eq:FP_b}) at incremental values of $L$ while maintaining constant values of all other parameters (see Appendix \ref{sec:SIMethods} for methods). Stringent tolerances were used to ensure the accurate detection of steady-states, and the amplitudes of inhibitor oscillations were compared over a significant timescale to extract limit cycle behaviour. Our analysis consists of a range of approximately $5 \times 10^4$ systems that exhibit undamped oscillations.
    
\begin{figure}[t!]
        \centering
            \includegraphics[width = 0.55\linewidth]{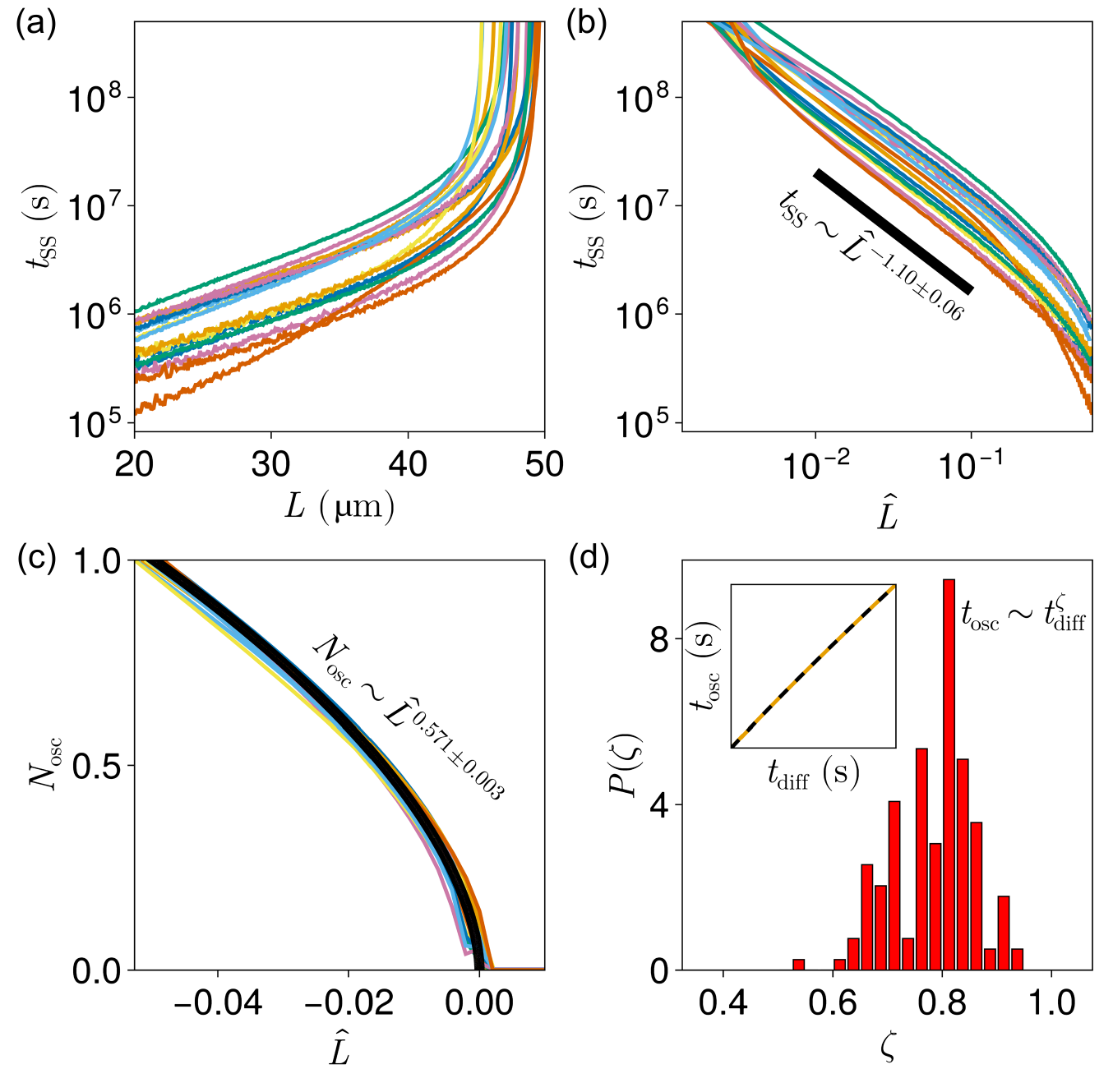}
            \caption[Activator-inhibitor systems exhibit critical behaviour]{\textbf{Activator-inhibitor systems exhibit critical behaviour.} (a,b) The time to reach steady-state for $20$ representative systems with $t_\text{SS}(L_\text{crit}) > 5 \times 10^8$ s and $45$ \textmu m $\leq L_\text{crit} \leq$ $50$ \textmu m, plotted as a function of: (a) system size less than the critical length; and (b) the relative system size $\hat{L}$ for $\hat{L} \geq 0$. Each colour represents a different system. The black line represents the average gradient obtained by fitting data within $10^{-2} \leq \hat{L} \leq 10^{-1}$ for all systems with $t_\text{SS}(L_\text{crit}) > 5 \times 10^8$ s using the equation $\log(t_\text{SS}) = \gamma \, \log(\hat{L}) + \log(t_0)$. (c) The normalised oscillation amplitude $N_\text{osc}$ (see Appendix \ref{sec:SIMethods} for methods) for the same systems as in (a,b) plotted against the relative system size $\hat{L}$. The black line represents the average profile obtained by fitting the plotted data for $\hat{L} \leq 0$ using the equation $N_\text{osc} = N_0 \, \hat{L} ^ \xi$. (d) Distribution of fit parameters $\zeta$ obtained by fitting the profiles of the oscillation timescale $t_\text{osc}$ as a function of the diffusive timescale $t_\text{diff} = ((L-w_A-w_I)^2 / 2) \, [ (1 / D_A) + (1 / D_E)]$ using the equation $t_\text{osc} = t_0 \, t_\text{diff}^\zeta$ for all systems with $t_\text{SS}(L_\text{crit}) > 5 \times 10^8$ s and $45$ \textmu m $\leq L_\text{crit} \leq$ $50$ \textmu m. Inset shows a representative fit (black, dashed) to the data (orange) for a single system shown in (a-c).}
            \label{fig:DivergingTimes}
    \end{figure}

    Notably, we observed a transition from damped to undamped oscillations when the system size exceeded a critical length $L = L_\text{crit}$. For example, the three systems with spatiotemporal oscillations plotted in Fig.~(\ref{fig:Oscillations}) have the same set of dynamical parameters but different system sizes, corresponding to: (left column) $L = L_\text{min} \ll L_\text{crit}$; (middle column) $L_\text{min} < L < L_\text{crit}$; and (right column) $L \gtrsim L_\text{crit}$. We found that the critical length is different for each system, can vary over a wide range of length-scales, and coincides with the system size at which the system's time to reach steady-state diverges (Fig.~(\ref{fig:DivergingTimes}a)). This latter point is related to the concept of critical slowing down, which occurs when eigenvalues approach zero and the system takes an increasingly long time to recover from perturbations \cite{zhang2022fgf8, dalwadi2023universal}.

    We can use the critical length to define a relative measure of how close a system is to exhibiting undamped oscillations, equal to $\hat{L} = (L_\text{crit} - L) / L_\text{crit}$, which is positive when $L < L_\text{crit}$, negative when $L > L_\text{crit}$, and zero at the critical point. When plotted as a function of $\hat{L}$, the profiles of the time taken to reach steady-state for each system collapse onto the same gradient (Fig.~(\ref{fig:DivergingTimes}a,b)). For eqs.~(\ref{eq:FP_a},\ref{eq:FP_b}) the mean critical exponent is $\gamma \simeq -1.1$. This finding of universal relative system dynamics near the critical point suggests a single shared mechanism of instability that is independent of the exact values of the dynamical parameters describing microscopic activator-inhibitor interactions.

    In addition, we see that profiles of the normalised oscillation amplitude plotted as a function of $\hat{L}$ collapse onto a single curve, revealing that each system exhibits similar oscillatory dynamics near the critical length (Fig.~(\ref{fig:DivergingTimes}c)). Importantly, the behaviour of the oscillation amplitude $N_\text{osc}$ normalised with respect to the final value at $L = L_\text{fin}$ follows the characteristic behaviour of a Hopf bifurcation near the critical point, corresponding to $N_\text{osc} \sim \hat{L}^{1/2}$ (see Appendix \ref{sec:SIMethods} for methods). We will elucidate this result further in the following section.

    A prediction that follows from our hypothesis that intrinsic diffusive delays can drive system instability is that perturbations in the activator and inhibitor levels must travel to regions of space where they more strongly influence system dynamics in order to induce a significant response. If this were not the case, interactions would occur instantaneously without any associated delay. This prediction implies that the oscillation period of the spatiotemporal oscillations presented in Fig.~(\ref{fig:Oscillations}) will be strongly correlated with the time it takes for the activator and inhibitor molecules to travel between these regions. We define the degree to which activator or inhibitor dynamics are influenced by small perturbations in the concentration of the other species as a `sensitivity' associated with the feedback. In the case of the activator-inhibitor feedback defined in eqs.~(\ref{eq:FP_a},\ref{eq:FP_b}), the sensitivity of the activator production rate to small perturbations in the inhibitor concentration, $\Sigma_A$, and the sensitivity of the inhibitor degradation rate to small perturbations in the activator concentration, $\Sigma_I$, can be quantified using the equations,
    \begin{align}
        \Sigma_A &= \left\lvert \partial_I \left[ \nu_A \left( \frac{\beta^h}{\beta^h + I^h} \right) \right] \right\rvert = \left\lvert J_{1,2} \right\rvert, \label{eq:Sensitivities_a} \\
        \Sigma_I &= \left\lvert \partial_A \left[ - k_I \left( \frac{\gamma}{\gamma + A} \right) I \right] \right\rvert = \left\lvert J_{2,1} \right\rvert, \label{eq:Sensitivities_b}
    \end{align}
    \noindent where $J_{i,j}$ are the components of the Jacobian defined in eq.~(\ref{eq:Jacobian}). The locations of maximum sensitivity can further be derived by differentiating eqs.~(\ref{eq:Sensitivities_a},\ref{eq:Sensitivities_b}), such that the maximum of $\Sigma_A$ occurs when $I = \beta \sqrt{(h-1)/(h+1)}$, and $\Sigma_I$ exhibits an asymptote towards positive infinity when $A \rightarrow -\gamma^{+}$ or negative infinity when $A \rightarrow -\gamma^{-}$. Since $A$ and $\gamma$ are strictly positive, the maximum observable value of $\Sigma_I$ corresponds to when $A = 0$.

    In the context of spatiotemporal oscillations, this result implies that a perturbation in the activator originating near its source must diffuse to where $I \simeq \beta$ in the limit $h \gg 1$ before it significantly influences inhibitor dynamics. By definition, this coincides with the edge of the activator source region. Similarly, the resulting perturbation in the inhibitor must diffuse to where $A$ is at its minimum value before it significantly influences activator dynamics. This results in the pair of intrinsic signalling delays $t_A \simeq (L-w_A-w_I)^2 / 2\,D_A$ and $t_I \simeq (L-w_A-w_I)^2 / 2\,D_I$, where $w_A \simeq \int_x (\beta^h / (\beta^h + I^h)) \, dx$, corresponding to the travel times of diffusing molecules between regions of maximum sensitivity. Notably, we expect from eqs.~(\ref{eq:Sensitivities_a},\ref{eq:Sensitivities_b}) that the total signalling delay $t_\text{diff} = t_A + t_I$ will be a good approximation of the oscillation period only when the off-diagonal elements of the Jacobian governing system dynamics are highly position dependent. In this limit, interactions occur only at point-like, spatially-separated interaction sites, since sensitivity to perturbations is minimal between these sites, and so intrinsic delays dominate system behaviour. In contrast, in systems where $J_{1,2}$ and $J_{2,1}$ do not depend on position then interactions occur locally without any intrinsic delays.    
    
    We tested this result by comparing the total signalling delay $t_\text{diff}$ to the oscillation period $t_\text{osc}$ while increasing the system size (Fig.~(\ref{fig:DivergingTimes}d)). Fitting the profiles of $t_\text{osc}$ for the $157$ systems with $t_\text{SS}(L_\text{crit}) > 5 \times 10^8$ s and $45$ \textmu m $\leq L_\text{crit} \leq$ $50$ \textmu m presented in Fig.~(\ref{fig:DivergingTimes}a-c) as a function of $t_\text{diff}$ with the function $t_\text{osc} = t_0 \, t_\text{diff}^\zeta$ reveals strong correlation between the two timescales, corresponding to an average value of $\zeta = 0.79 \pm 0.07$. This correlation implies that the oscillation period is constrained by the diffusive dynamics of activator and inhibitor molecules, but we predict that deviation from $\zeta = 1$ is the result of additional contributions to the oscillation timescale from the reaction terms of eqs.~(\ref{eq:FP_a},\ref{eq:FP_b}). Together, these results reinforce our hypothesis that spatiotemporal oscillations, and ultimately a transition from damped to undamped oscillations, are driven by intrinsic diffusive delays.

    \subsection{Investigating the Hopf bifurcation} \label{sec:NumericalResults}

    We next investigate the nature of the transition from damped to undamped oscillations by numerically solving eqs.~(\ref{eq:FP_a},\ref{eq:FP_b}) for spatially heterogeneous base states and determining the corresponding eigenvalues. This numerical analysis allows us to access the base states corresponding to unstable fixed points beyond the critical length, which are not accessible using simulations where small perturbations at system sizes $L \geq L_\text{crit}$ drive the system towards a limit cycle as shown in the right column of Fig.~(\ref{fig:Oscillations}). In this section, we implement a discretised form of the Jacobian defined in eq.~(\ref{eq:Jacobian}) to analyse system stability for spatially heterogeneous base states across a wide range of system sizes.

    \begin{figure}[t!]
        \centering
            \includegraphics[width = 0.85\linewidth]{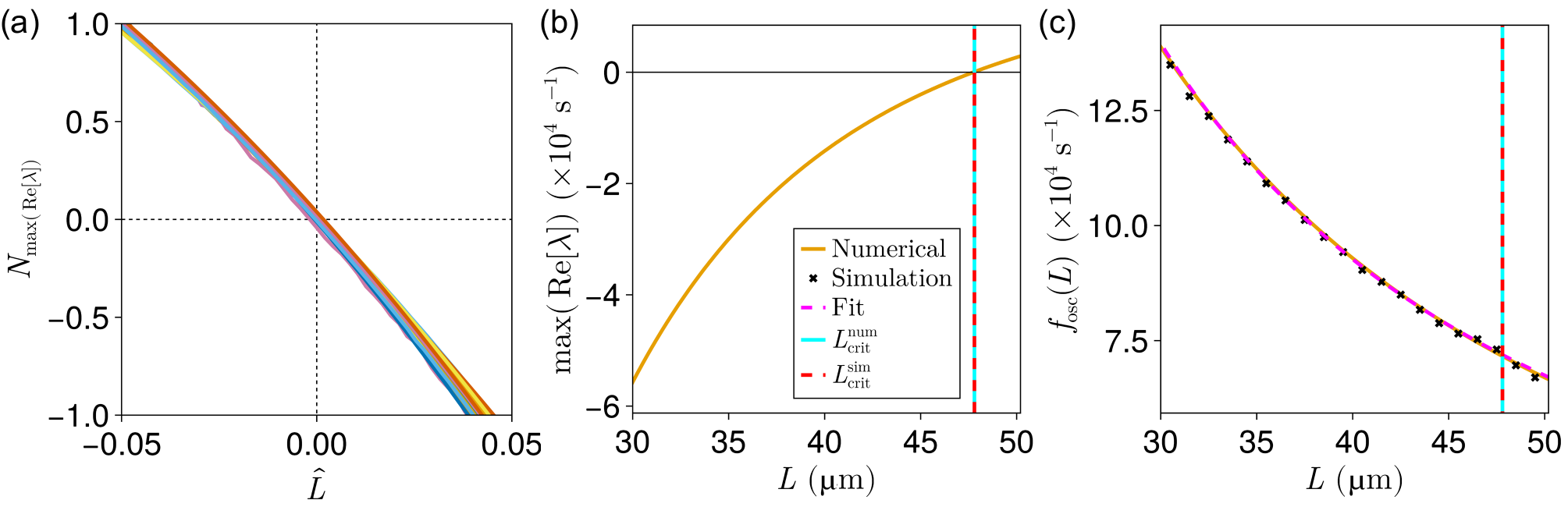}
            \caption[The transition from damped to undamped oscillations coincides with a Hopf bifurcation]{\textbf{The transition from damped to undamped oscillations coincides with a Hopf bifurcation.} (a) The normalised real part of the eigenvalue with maximum real part $N_{\max(\text{Re}[\lambda])}$ (see Appendix \ref{sec:SIMethods} for methods) for the $20$ representative systems shown in Fig.~(\ref{fig:DivergingTimes}a-c) with $t_\text{SS}(L_\text{crit}) > 5 \times 10^8$ s and $45$ \textmu m $\leq L_\text{crit} \leq$ $50$ \textmu m, plotted as a function of the relative system size $\hat{L}$. Each colour represents a different system. The black dashed lines represent the axes where $\hat{L} = 0$ (corresponding to $L = L_\text{crit}$) or $N_{\max(\text{Re}[\lambda])}=0$. (b,c) For a single representative system, values of: (b) the real part; and (c) the imaginary part (equal to the oscillation frequency $f_\text{osc}$), of the eigenvalue with maximum real part plotted as a function of increasing system size. Black crosses in (c) indicate oscillation frequencies obtained directly from simulations (see Appendix \ref{sec:SIMethods} for methods), and the magenta, dashed line represents the fit $f_\text{osc}(L) = r \, L^{-s}$ applied to the numerical data. The critical lengths obtained numerically (red) and using simulations (cyan, dashed) perfectly overlap. The key for (b) and (c) is shown in (b). System parameters are presented in tables in Appendix \ref{sec:SIMethods}.}
            \label{fig:NumericalHopf}
    \end{figure}

    Our numerical solutions at stable fixed points ($L < L_\text{crit}$) successfully recapitulated the results of our simulations. Using our numerical framework, we demonstrated that the transition from damped to undamped oscillations in the ER model corresponds to a Hopf bifurcation: at the critical length an eigenvalue emerges with a positive real part and non-zero imaginary part (Fig.~(\ref{fig:NumericalHopf})). We also show that the evolution of the numerical oscillation frequencies, extracted from the imaginary components of the eigenvalues with maximum real parts, can be well fit by the function $r \, L^{-s}$ (Fig.~(\ref{fig:NumericalHopf}c)). Assuming that $w_A, w_I \sim L$, such that $t_\text{diff} \sim L^2$, the fitted value of $b=1.4232 \pm 0.0013$ corresponds to $f_\text{osc} \sim t_\text{diff}^{-0.71}$, which is similar to the value of $\zeta = 0.79 \pm 0.07$ linking the oscillation and diffusion timescales obtained through simulations Fig.~(\ref{fig:DivergingTimes}d). Once again, this highlights the role of intrinsic diffusive delays in the maintenance of spatiotemporal oscillations.

\begin{figure}[t!]
        \centering
            \includegraphics[width = 0.88\linewidth]{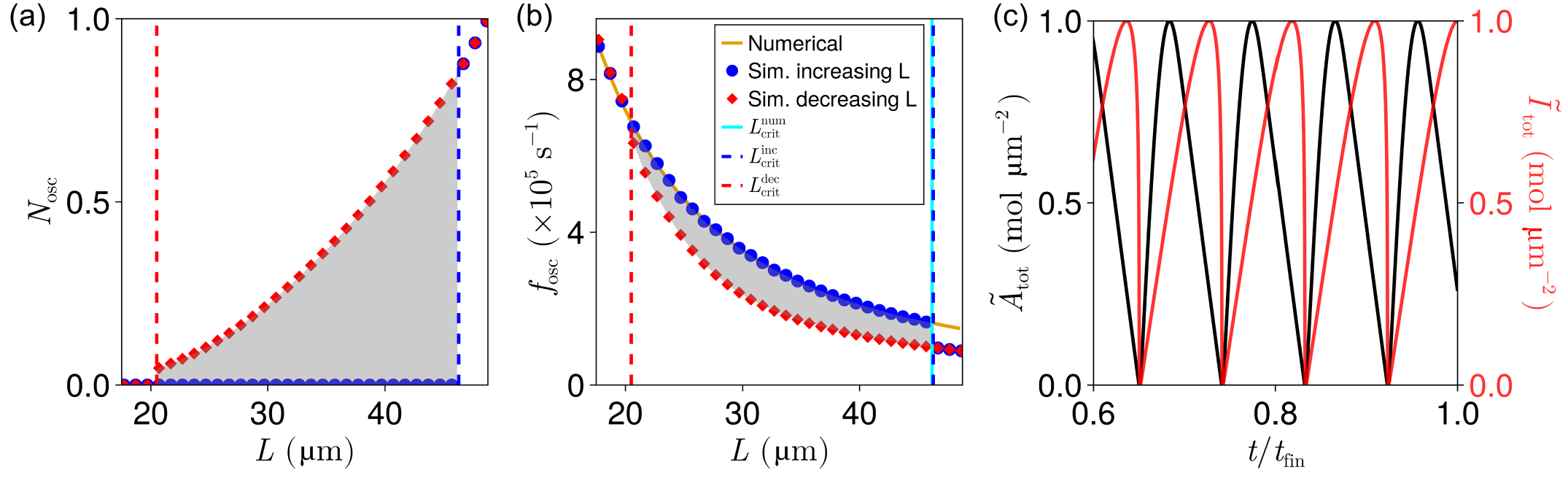}
            \caption[Systems can exhibit sub-critical Hopf bifurcations]{\textbf{Systems can exhibit sub-critical Hopf bifurcations.} Measurements of (a) the normalised oscillation amplitude $N_\text{osc}$ (see Appendix \ref{sec:SIMethods} for methods); and (b) the oscillation frequency, as functions of system size $L$, for simulations where $L$ is increased from $L_\text{init}$ (blue circles) and $L$ is decreased from $L_\text{crit}$ (red diamonds). The blue and red dashed lines are the corresponding critical lengths that mark the separation between damped and undamped oscillations. The grey shaded region represents the difference between the normalised amplitudes measured when increasing or decreasing $L$. In (b) the imaginary part of the eigenvalue with maximum real part obtained numerically is also plotted, and is independent of whether $L$ is increasing or decreasing. The critical lengths obtained numerically (cyan) and using simulations with increasing $L$ (blue, dashed) perfectly overlap. The key for (a,b) is shown in (b). (c) The total activator ($A_\text{tot} = \int_x A \, dx$, $\tilde{A}_\text{tot} = (A_\text{tot} - \min(A_\text{tot})) / (\max(A_\text{tot}) - \min(A_\text{tot}))$ where max and min correspond to data within the plotted time window) and inhibitor (calculated using the same method as for $A_\text{tot}$) concentrations for a representative system at system size $L = 48.7$ \textmu m obtained by simulating eqs.~(\ref{eq:FP_a},\ref{eq:FP_b}), plotted as a function of fractional simulation time $t/t_\text{fin}$, where $t_\text{fin}$ is the termination time of the simulation and varies between simulations. System parameters are presented in tables in Appendix \ref{sec:SIMethods}.}
            \label{fig:Hysteresis}
    \end{figure}

    In addition to the Hopf bifurcation, we observed a minority of systems that exhibited large changes in their oscillation amplitudes and frequencies at the critical system size $L_\text{crit}$ in simulations (Fig.~(\ref{fig:Hysteresis}a,b)), which we attribute to sub-critical Hopf bifurcations. These systems were identified by comparing the oscillation frequencies extracted from simulations and obtained numerically (see Appendix \ref{sec:SIMethods} for methods). This is possible since the larger amplitude oscillations observed in simulations correspond to dynamics far from the unstable fixed point used to calculate the numerical oscillation frequency, and the nonlinear effects assumed to be negligible in the calculations of eigenvalues influence only the oscillation frequency observed in simulations (Fig.~(\ref{fig:Hysteresis}b)). Additional evidence of these nonlinear effects can be observed through the asymmetry between activator and inhibitor oscillations (compare Fig.~(\ref{fig:Hysteresis}c) to Fig.~(\ref{fig:Oscillations}a)), and the loss of the characteristic behaviour $N_\text{osc} \sim \hat{L}^{1/2}$ (compare Fig.~(\ref{fig:Hysteresis}a) to Fig.~(\ref{fig:DivergingTimes}c)). Notably, simulations with these properties also demonstrate hysteresis effects, whereby the critical length changes depending on whether the system was growing or shrinking (Fig.~(\ref{fig:Hysteresis}a,b)). 

    Together, our results reveal that an activator-inhibitor system that can be described by eqs.~(\ref{eq:FP_a},\ref{eq:FP_b}) can undergo a Hopf bifurcation when its system size exceeds a critical length. Coupled with the results presented in sections \ref{sec:ModelSystem} \& \ref{sec:SimResults}, we propose that this bifurcation is the result of intrinsic diffusion-driven delays that require spatial heterogeneity in (at least) the definition of the inhibitor source region.

    \subsection{Simple DDE model}

    Due to the analytical intractability of the coupled PDEs defined in eqs.~(\ref{eq:FP_a},\ref{eq:FP_b}), we introduce a tractable DDE model based on eqs.~(\ref{eq:FP_a},\ref{eq:FP_b}) with the goal of understanding which system parameters are important for causing a Hopf bifurcation. Specifically, in this section we test whether the same oscillatory behaviour and bifurcation exhibited by the full system can be derived analytically for an equivalent ODE system where intrisinc diffusive delays are introduced explicitly.
    
    We assume that there exists $0$D pools of activator and inhibitor species that interact purely through delayed feedback, approximating the limit of eqs.~(\ref{eq:FP_a},\ref{eq:FP_b}) where the spatially heterogeneous sensitivities for activator feedback on the inhibitor and vice-versa are high at a single position and zero everywhere else. We additionally assume that these positions of high sensitivity are near the activator and inhibitor sources, where the diffusion terms of eqs.~(\ref{eq:FP_a},\ref{eq:FP_b}) do not strongly affect their steady-state concentrations. In this case, the diffusivities enter the DDEs only through the delays $t_A$ and $t_I$, representing the time it takes for activator or inhibitor molecules to travel between the positions of high sensitivity. Using these assumptions, we define the non-dimensionalised, $0$D DDEs for this simplified system as,
    \begin{align}
        \dot{X}(\tau) &=  \left( \frac{\sigma \, b^h}{b^h + Y(\tau - \tau_I)^h} \right) - \left( \frac{\sigma \, a^m \, X(\tau)}{a^m + X(\tau)^m} \right), \label{eq:DDE_a} \\
        \dot{Y}(\tau) &= 1 - \left( \frac{g^n \, Y(\tau)}{g^n + X(\tau - \tau_A)^n} \right), \label{eq:DDE_b}
    \end{align}
    \noindent where overdots represent time derivatives, $X = A/(\nu_A / k_A)$, $Y = I/(\nu_I / k_I)$, $a = \alpha/(\nu_A / k_A)$, $b = \beta/(\nu_I / k_I)$, $g = \gamma/(\nu_A / k_A)$, $\sigma = k_A / k_I$, $\tau = k_I \, t$, and $\tau_{A,I} = k_I \, t_{M,E}$. Notably, we have added the constants $n$ and $m$ to eqs.~(\ref{eq:DDE_a},\ref{eq:DDE_b}) to generalise the form of the activator-inhibitor feedback, while the full system defined by eqs.~(\ref{eq:FP_a},\ref{eq:FP_b}) uses the values $n=m=1$.

\begin{figure}[t!]
        \centering
            \includegraphics[width = 0.6\linewidth]{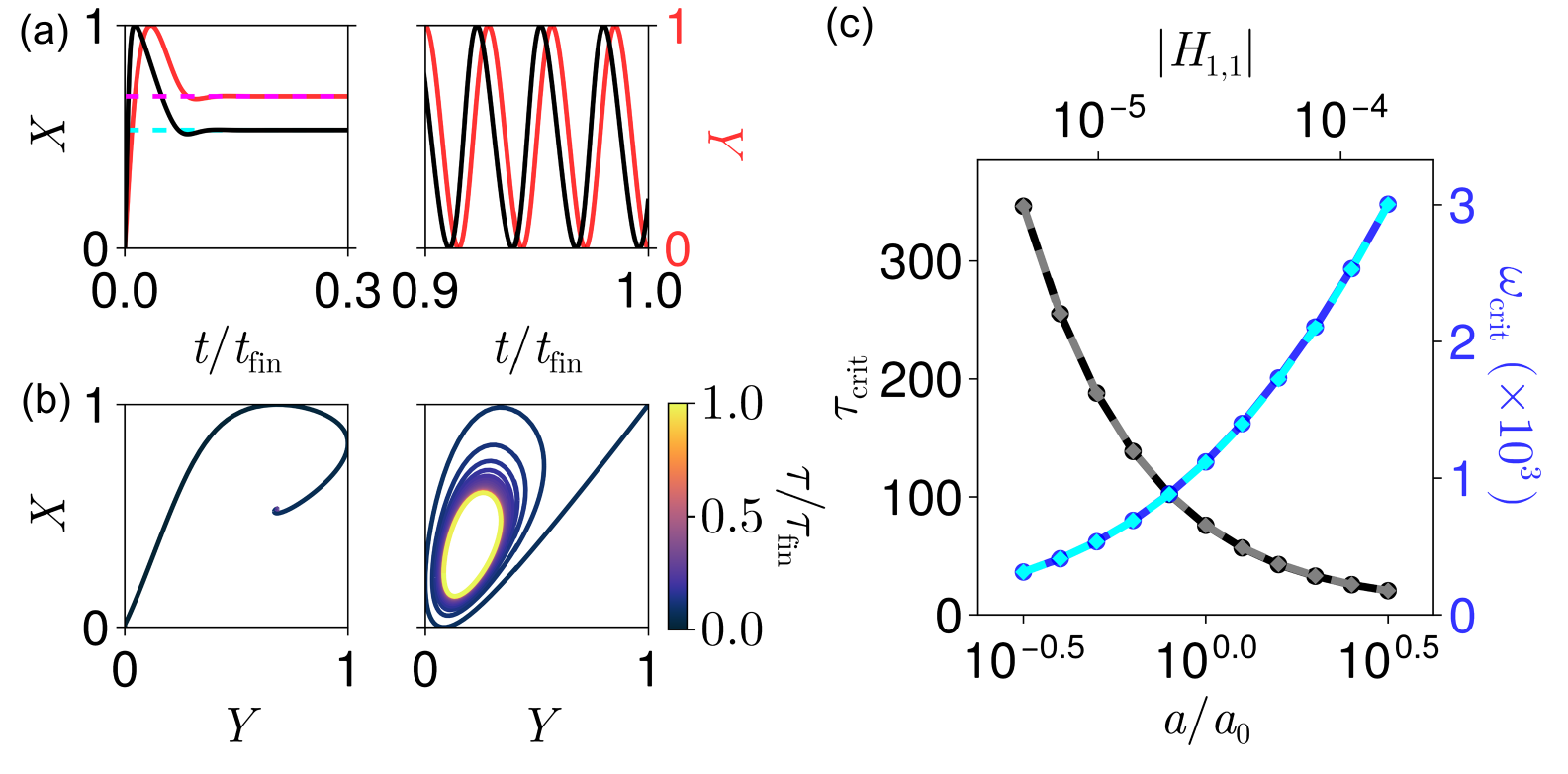}
            \caption[A simplified delay model exhibits Hopf bifurcations]{\textbf{A simplified delay model exhibits Hopf bifurcations.} (a) The activator ($X$) and inhibitor ($Y$) concentrations obtained by simulating eqs.~(\ref{eq:DDE_a},\ref{eq:DDE_b}), plotted as a function of fractional simulation time $\tau/\tau_\text{fin}$, where $\tau_\text{fin}$ is the termination time of the simulation and varies between simulations. The cyan and magenta dashed lines in correspond to the numerical solutions of eqs.~(\ref{eq:DDE_a},\ref{eq:DDE_b}) for the activator and inhibitor concentrations, respectively. (b) The corresponding data plotted in concentration-space, where the colour indicates the fractional simulation time. (c) For constant values of $b$, $g$ and $\sigma$, the values of the critical delay and oscillation frequency at the critical point defined in eq.~(\ref{eq:AnalyticalThreshold1},\ref{eq:AnalyticalThreshold2}), and the equivalent values extracted from simulations of eqs.~(\ref{eq:DDE_a},\ref{eq:DDE_b}), plotted as a function of varying $a$, approximately equivalent to varying $|H_{1,1}|$. Analytical data (diamond points, dashed lines) agree well with simulation data (circle points, solid lines). System parameters are presented in tables in Appendix \ref{sec:SIMethods}.}
            \label{fig:DDE}
    \end{figure}

    We first demonstrate that eqs.~(\ref{eq:DDE_a},\ref{eq:DDE_b}) can reproduce the oscillatory behaviour exhibited by the full model using simulations (see Appendix \ref{sec:SIMethods} for methods). We observe strong qualitative similarities between the dynamics displayed by the full model in Fig.~(\ref{fig:Oscillations}) and the DDE model in Fig.~(\ref{fig:DDE}a,b).

    In order to understand the oscillatory dynamics of the simplified system, we perform a stability analysis around the steady-state $\bar{u} = (\bar{X}, \bar{Y})$ with small perturbations $\underline{\delta u} = (\delta X, \delta Y) = \sum_i \underline{v}_i \exp(\lambda_i \, \tau)$. The delayed feedback in eqs.~(\ref{eq:DDE_a},\ref{eq:DDE_b}) prevents them being able to be written in matrix form following linearisation, and instead we define an effective Jacobian $\underline{\underline{H}}$ that corresponds to $\partial_\tau \, \underline{\delta u} \simeq \underline{\underline{H}} \, \underline{\delta u}$ in the limit that the delays $\tau_{A,I} \rightarrow 0$ (see Appendix \ref{sec:SI2}),
    
    \begin{equation}
        \underline{\underline{H}} = \begin{bmatrix}
            \sigma \, a^m \left( \frac{(m-1) \bar{X}^m - a^m}{\left( a^m + \bar{X}^m \right)^2} \right) & - \sigma \left( \frac{h \, b^h \, \bar{Y}^{h-1}}{\left( b^h + \bar{Y}^h \right)^2} \right) \\
            \left( \frac{n \, g^n \, \bar{X}^{n-1}}{\left( g^n + \bar{X}^n \right)^2} \right) & - \left( \frac{g^n}{g^n + \bar{X}^n} \right)
        \end{bmatrix}.
        \label{eq:DDEJacobianMainText}
    \end{equation}
    
    We seek conditions where the perturbations $\delta X$ and $\delta Y$ grow with time, or where a value of $\lambda_i$ first exhibits a real part greater than zero. We will denote this eigenvalue $\Lambda$, and values of $\Lambda = 0$ and $\Lambda = i \, \omega$ correspond to the onset of non-oscillatory (saddle-node, pitchfork etc.) and Hopf bifurcations, respectively. Defining $H_{i,j}$ as the components of the effective Jacobian defined in eq.~(\ref{eq:DDEJacobianMainText}), we extract the pair of characteristic equations (see Appendix \ref{sec:SI2}),
    \begin{align}
        \Lambda - H_{1,1} - \left( \frac{V_2}{V_1} \right) H_{1,2} \, \exp(- \Lambda \, \tau_I) &= 0, \label{eq:Characteristic_a} \\
        \qquad \Lambda - H_{2,2} - \left( \frac{V_1}{V_2} \right) H_{2,1} \, \exp(- \Lambda \, \tau_A) &= 0, \label{eq:Characteristic_b}
    \end{align}
    \noindent where $\underline{V}=(V_1,V_2)$ represents the vector $\underline{v}_i$ corresponding to the eigenvalue $\Lambda$.    
    
    Using eqs.~(\ref{eq:Characteristic_a},\ref{eq:Characteristic_b}), we can determine analytically that when $m=n=1$, corresponding to the feedback in eqs.~(\ref{eq:FP_a},\ref{eq:FP_b}), non-oscillatory bifurcations are only possible when $\text{sgn}(h) < 0$ (see Appendix \ref{sec:SI2}). This is not true for the simulations of the full system shown in Figs.~(\ref{fig:Oscillations}-\ref{fig:Hysteresis}), which use the value of $h=4$ (see Appendix \ref{sec:SIMethods} for system parameters used in simulations). We are also able to derive that Hopf bifurcations arise for the feedback defined in eqs.~(\ref{eq:FP_a},\ref{eq:FP_b}) at values of the total diffusive delay $\tau_A + \tau_I$ and the corresponding oscillation frequency $\omega$ equal to (see Appendix \ref{sec:SI2}),
    \begin{equation}
        \tau_A + \tau_I = \left( \frac{1}{\omega} \right) \tan^{-1} \left( \frac{ \omega \left( |H_{1,1}| + |H_{2,2}| \right)}{ \omega^2 - |H_{1,1}||H_{2,2}|} \right), \label{eq:AnalyticalThreshold1}
    \end{equation}
    \begin{equation}
        \omega = \left\{ -\left( \frac{|H_{1,1}|^2 + |H_{2,2}|^2}{2} \right) + \left[ \left( \frac{|H_{1,1}|^2 - |H_{2,2}|^2}{2} \right)^2 + |H_{1,2}|^2 |H_{2,1}|^2 \right]^{\,\frac{1}{2}} \right\}^{\frac{1}{2}}. \label{eq:AnalyticalThreshold2}
    \end{equation}
    \noindent More specifically, we can deduce that undamped oscillations arise when $\tau_A + \tau_I$ exceeds the threshold defined in eq.~(\ref{eq:AnalyticalThreshold1}) and $\omega$ is less than the value at the critical point defined in eq.~(\ref{eq:AnalyticalThreshold2}). From eq.~(\ref{eq:AnalyticalThreshold1}), the final requirements that the critical value of the delay satisfies $\tau_A + \tau_I > 0$ and the critical value of the oscillation frequency is real set that $\omega^2 > |H_{1,1}| |H_{2,2}|$ and $|H_{1,2}| |H_{2,1}| > |H_{1,1}| |H_{2,2}|$, respectively, in order to observe a Hopf bifurcation. For the remainder of our analysis we assume that $\tau_A=\tau_I$.

    We find that eqs.~(\ref{eq:AnalyticalThreshold1},\ref{eq:AnalyticalThreshold2}) correctly predict the critical value of the delay $\tau_\text{crit}$ and the oscillation frequency at the critical point $\omega_\text{crit}$ (Fig.~(\ref{fig:DDE}c)). We observe in Fig.~(\ref{fig:DDE}c) that increasing $a$ reduces the stability of the system by increasing its susceptibility to a Hopf bifurcation at shorter delays. We hypothesise that this is the result of increasing $a$ ultimately increasing the self-repressive effect of $X$ (increasing the magnitude of $H_{1,1}$), which is important for driving system oscillations. This could correspond to a novel route for improving the stability of the full system: by reducing the self-repression of the activator.
    
    Overall, we find that explicitly imposing delays in a simplified toy model can recreate the oscillations and bifurcations arising from intrinsic signalling delays in the full model. While the explicitly imposed delay does not provide new information about the mechanistic origin of intrinsic delays, it paves the way for a two-pronged analysis where key drivers of oscillations can be identified analytically in DDEs before being probed in more complex spatial models.

    \section{Discussion}

    In this work we have investigated the origin and nature of spatiotemporal oscillations generated by an activator-inhibitor system with feedback through both degradation and production but with no explicit delays. Using a combination of simulations, numerical analysis and analytical derivations we have demonstrated that intrinsic diffusion-driven delays arising from spatially heterogeneous feedback can drive a transition from damped to undamped spatiotemporal oscillations for systems with spatially extended regions of interaction, and that this transition can be formally characterised as a Hopf bifurcation.
    
    We propose that the mechanism of intrinsic diffusion-driven delays generating spatiotemporal oscillations and ultimately a Hopf bifurcation derived in this work is a generalisable property of a wide range of biological and physical systems. In particular, we expect spatiotemporal oscillations to arise in general for systems where signalling is mediated by diffusive species, and for Hopf bifurcations to arise when these systems grow beyond a critical size. However, an additional important property of the activator-inhibitor feedback studied in this work is its ability to generate solutions that scale with system size \cite{mosby2026position}. This ensures that the activator and inhibitor can continue interacting as the system grows. Systems without this scaling property may exhibit critical system sizes beyond the limits where the activator and inhibitor can interact, preventing the onset of a Hopf bifurcation. The analytical and computational pipeline derived in this work can be used to probe the existence and nature of spatiotemporal oscillations in a general system with interactions mediated by diffusive species. In particular, it is of interest to compare the critical exponents extracted from the divergence of relaxation timescales for different systems to test whether universal rules exist that govern the onset of Hopf bifurcations in these systems.
    
    As predicted from dynamical systems theory, the Hopf bifurcation in our system is accompanied by diverging relaxation timescales (critical slowing down) and universal relative system dynamics \cite{simons2026cell}. It has been proposed that some biological systems operate in a stable regime near criticality to optimise specific system properties. For example, the mammallian cochlea operates near a Hopf bifurcation to improve amplitude amplification and frequency selectivity \cite{alonso2025amplification}, and differentiating cells operate near a saddle-node bifurcation to retain the ability to respond to inhibitory signals for a prolonged period before irreversibly transitioning to a cell fate \cite{simons2026cell}. The activator-inhibitor feedback studied in this work mirrors that of the expansion-repression (ER) model used to describe pattern scaling through morphogen-expander feedback \cite{ben2010scaling, mosby2026position}. We propose that the existence of a Hopf bifurcation could amplify the scaling properties of the ER model, since these bifurcations have been linked to system-wide co-ordination through long-range spatial and temporal correlations \cite{di2018landau}. Despite the potential advantages of activator-inhibitor feedback in biological systems, the results of this work suggest that additional `safety mechanisms' may be necessary to minimise the negative impacts of spatiotemporal oscillations in certain systems. For example, our findings could imply previously unanticipated functions for the growth arrest of biological systems with ER feedback \cite{martin2009cell, aguilar2018critical} and the irreversibility of some chemo-morphogenetic interactions \cite{autorino2025closed}: preventing the onset of a Hopf bifurcation.
    

    Finally, we note that although our analysis of the delay-differential equation (DDE) model reinforces our hypothesis that intrinsic diffusion-driven delays drive a Hopf bifurcation in systems with activator-inhibitor feedback, further mathematical analysis is required to concretely relate diffusive delays to wave-like solutions in our model.

    \clearpage

    \setcounter{section}{0}
    \setcounter{table}{0}
    \setcounter{figure}{0}
    \setcounter{equation}{0}

    \renewcommand{\thesection}{A\arabic{section}}
    \renewcommand{\thetable}{A\arabic{table}}
    \renewcommand{\thefigure}{A\arabic{figure}}
    \renewcommand{\theequation}{A\arabic{equation}}

    \section{Methods} \label{sec:SIMethods}

    \subsection{Locating systems with undamped oscillations} \label{sec:FindingSysMethods}

    We extracted an initial sample of systems that exhibited undamped oscillations at system size $L=50$ \textmu m by simulating eqs.~(\ref{eq:FP_a},\ref{eq:FP_b}) and sweeping over the parameters $D_{A,I}$, $k_{A,I}$ and $\nu_{A,I}$. For this preliminary sweep we defined undamped oscillations using the algorithm presented in our previous work \cite{mosby2026position}. This means that in this work we only analyse systems with $L_\text{crit} \leq 50$ \textmu m, although we expect the results to be general for any value of $L_\text{crit}$.

    Our preliminary sweep consisted of $898 \, 685$ simulations, of which $64 \, 133$ exhibited undamped oscillations at $L=50$ \textmu m. In this work we analysed a subset of $46 \, 499$ of these systems, of which $147$ were found to exhibit hysteresis effects. We identified these systems as those with oscillation frequencies extracted from simulations ($f_\text{osc}^\text{sim}$) and obtained numerically ($f_\text{osc}^\text{num}$) that satisfy the conditions: (i) $f_\text{osc}^\text{sim}(L_\text{crit} - \Delta L) < 1.05 \, f_\text{osc}^\text{num}$; (ii) $f_\text{osc}^\text{sim}(L_\text{crit}) > f_\text{osc}^\text{num}(L_\text{crit}) / 1.05$; and (iii) $f_\text{osc}^\text{sim}(L_\text{crit}) > 1.25 \, f_\text{osc}^\text{num}$ or $f_\text{osc}^\text{sim}(L_\text{crit}) < f_\text{osc}^\text{num}(L_\text{crit}) / 1.25$. These conditions capture systems with significant deviations in the oscillation frequency extracted from simulations at the critical point that are not present in the numerical measurements.

    \subsection{Simulation methods}

    Eqs.~(\ref{eq:FP_a},\ref{eq:FP_b}) in the main text were simulated using the Rodas4P solver implemented in the Julia programming language. This solver utilises adaptive time-stepping to significantly reduce run-time. When plotting the time-dependent evolution of the activator and inhibitor levels in Fig.~(\ref{fig:Oscillations}), simulations were repeated and data was extracted at the intervals $10^{\lfloor \log(t_\text{fin} / 10^5) \rfloor}$, where $t_\text{fin}$ is the original simulation run-time and $\lfloor \cdots \rfloor$ represents the floor function. Simulations were carried out on a spatial grid of $N=201$ points. Simulations initially used a domain of size $L_\text{min}=0.05 L_\text{max}$, where $L_\text{max} = 100$ \textmu m, with initial conditions $A(x,t=0)=I(x,t=0)=0$ $\forall x$. In subsequent simulations the domain size was increased incrementally by $\Delta L = 0.001 L_\text{max}$. The steady-state activator and inhibitor profiles were used as initial conditions for the next simulation with system size $L+\Delta L$. The parameters used to generate Figs.~(\ref{fig:Oscillations},\ref{fig:NumericalHopf}b,c,\ref{fig:Hysteresis},\ref{fig:DDE}) are presented in tables \ref{tab:GlobalParameters} \& \ref{tab:IndividualParameters}.
    
    Simulations with system sizes less than the critical length were terminated using a bespoke algorithm that tests for steady-state by comparing the changes in activator and inhibitor levels between timesteps to relative ($10^{-6}$) and absolute ($10^{-8}$) tolerances over a prolonged period of time. This was implemented to ensure that simulations were not terminated due to slow modulation of activator or inhibitor levels near a maximum or minimum. Undamped oscillations were detected using a more stringent version of the algorithm presented in our previous work \cite{mosby2026position}. Simulations were terminated if the total inhibitor concentration integrated over the spatial domain of the system ($I_{\text{tot}}(t) = \int_x I \, dx$) satisfied the conditions,
    \begin{equation}
        \vline \,\, \min \left( \frac{I_{\text{tot}}(t_\text{max}^{(i)}) - I_{\text{tot}}(t_\text{min}^{(i)})}{I_{\text{tot}}(t_\text{max}^{(j)}) - I_{\text{tot}}(t_\text{min}^{(j)})}, \frac{I_{\text{tot}}(t_\text{max}^{(j)}) - I_{\text{tot}}(t_\text{min}^{(j)})}{I_{\text{tot}}(t_\text{max}^{(i)}) - I_{\text{tot}}(t_\text{min}^{(i)})} \right) - 1 \,\, \vline < 0.005 \qquad \text{ for any } j \leq i - 10,
        \label{eq:OscillationCheck}
    \end{equation}
    \noindent where the times $t_\text{min,max}^{(i)}$ correspond to the $i^\text{th}$ occurrence of when $\partial_t \, I_{\text{tot}}(t) = 0$, and either $\partial_t^2 \, I_{\text{tot}}(t) > 0$ for the minima times $t_\text{min}^{(i)}$ or $\partial_t^2 \, I_{\text{tot}}(t) < 0$ for the maxima times $t_\text{max}^{(i)}$. The total activator level can be used in eq.~(\ref{eq:OscillationCheck}) instead, but only the dynamics of one species needs to be checked to verify the presence of undamped oscillations.
    
    We included additional constraints that caused the arrays containing the maxima and minima of $I_{\text{tot}}$ to be reset. One of these constraints was that the extrema could not exhibit net translation, which was defined as either: $I_{\text{tot}}(t_\text{max}^{(i)}) - I_{\text{tot}}(t_\text{max}^{(i-10)}) > 0.001$ and $I_{\text{tot}}(t_\text{min}^{(i)}) - I_{\text{tot}}(t_\text{min}^{(i-10)}) > 0.001$; or $I_{\text{tot}}(t_\text{max}^{(i)}) - I_{\text{tot}}(t_\text{max}^{(i-10)}) < -0.001$ and $I_{\text{tot}}(t_\text{min}^{(i)}) - I_{\text{tot}}(t_\text{min}^{(i-10)}) < -0.001$. Another constraint was that the change in oscillation period could not vary significantly, which was defined as any of: $t_\text{max}^{(i)} - t_\text{max}^{(i-1)} > 1.1 (t_\text{max}^{(i-9)} - t_\text{max}^{(i-10)})$; $t_\text{max}^{(i)} - t_\text{max}^{(i-1)} < (t_\text{max}^{(i-9)} - t_\text{max}^{(i-10)}) / 1.1$; $t_\text{min}^{(i)} - t_\text{min}^{(i-1)} > 1.1 (t_\text{min}^{(i-9)} - t_\text{min}^{(i-10)})$; or $t_\text{min}^{(i)} - t_\text{min}^{(i-1)} < (t_\text{min}^{(i-9)} - t_\text{min}^{(i-10)}) / 1.1$. The final constraint was that the arrays were reset following an inflection point, which was defined as when the number of recorded maxima and minima escape the limits $N_\text{min} - 1 \leq N_\text{max} \leq N_\text{min} + 1$. Simulations were carried out while incrementally increasing systems size until $25$ successive simulations exhibited undamped oscillations. This results in the critical length satisfying $L_\text{crit} = L_\text{fin} - 24\Delta L$, where $L_\text{fin}$ is the final simulated system size before termination.

    Average oscillation frequencies were calculated using either the final $10$ maxima and minima of $I_{\text{tot}}$, or all of the observed maxima and minima if that number was less than $10$. We fitted the quadratic polynomial $a (t - t_0)^2 + b (t - t_0) + c$ to the $5$ points surrounding each $I_{\text{tot}}$ maxima and minima, and calculated the average oscillation frequency as the reciprocal of the average time between maxima, $1 / \langle t_0^{(i + 1)} - t_0^{(i)} \rangle$. Average oscillation amplitudes were extracted as the average change in the fit parameter $c$ between adjacent $I_{\text{tot}}$ maxima and minima. The normalised amplitudes $N_\text{osc}$ in Figs.~(\ref{fig:DivergingTimes}c \& \ref{fig:Hysteresis}a) are equal to the amplitudes $A_\text{osc}$ divided by the value obtained at $L=L_\text{fin}$.

    Eqs.~(\ref{eq:DDE_a},\ref{eq:DDE_b}) in the main text were simulated using the Tsit5 solver and the method of steps implemented in the Julia programming language. We used the same algorithms for detecting steady-state and undamped oscillations as in simulations of eqs.~(\ref{eq:FP_a},\ref{eq:FP_b}). In this case, the total activator and inhibitor levels are equivalent to the activator and inhibitor levels at the single effective position being studied.

    \subsection{Numerical methods}

    Eqs.~(\ref{eq:FP_a},\ref{eq:FP_b}) in the main text were solved numerically using the Newton Raphson method implemented in the Julia programming language. The spatial grid ($N=201$) and system sizes ($L_\text{min}=0.05 L_\text{max}$, $L_\text{max} = 100$ \textmu m, and $\Delta L = 0.001 L_\text{max}$) were identical to those used in simulations, and the relative and absolute tolerances of the solver were $10^{-6}$ and $10^{-8}$, respectively. At $L_\text{min}=0.05 L_\text{max}$, the steady-state activator and inhibitor levels obtained from simulations at this system size were used as the initial conditions for the numerical solver. Subsequently, the activator and inhibitor profiles obtained numerically at system size $L - \Delta L$ were used as the initial condition at system size $L$. This process was continued until $L = L_\text{fin}$, corresponding to the maximum system size used in corresponding simulations.

    Eigenvalues were obtained by expanding the Jacobian defined in eq.~(\ref{eq:Jacobian}) in the main text to take into account the discretised spatial grid. This Jacobian acts on a vector of length $2N$ of the form $(A_1, \cdots, A_{N}, I_1, \cdots, I_N)$. After numerically solving eqs.~(\ref{eq:FP_a},\ref{eq:FP_b}) in the main text, we calculate the eigenvalue of the Jacobian with a maximum real part using a partial Schur decomposition and the Arnoldi method (part of the ArnoldiMethod package in Julia). The normalised real part of the eigenvalue with maximum real part $N_{\max(\text{Re}[\lambda])}$ in Fig.~(\ref{fig:NumericalHopf}a) corresponds to $\max(\text{Re}[\lambda])$ divided by the corresponding value at $L=L_\text{fin}$.
    
    Eqs.~(\ref{eq:DDE_a},\ref{eq:DDE_b}) in the main text were solved numerically using the same Tsit5 solver and method of steps algorithm, but with the relative and absolute tolerances of $10^{-8}$ and $10^{-10}$, respectively. The resulting activator and inhibitor levels were compared to the steady-state outputs of simulations in the left column of Fig.~(\ref{fig:DDE}a).

    \begin{table}[h!]
        \centering
            \begin{tabular}{ x{2cm} x{3cm} }
				\textbf{Parameter} & \textbf{Value} \\
				\hline
			$\alpha$ & 1.000 mol \textmu m${}^{-2}$ \\
			$\beta$ & 0.500 mol \textmu m${}^{-2}$ \\
			$\gamma$ & 1.000 mol \textmu m${}^{-2}$ \\
            $h$ & 4 \\
			$w_I / L$ & 0.200 \\
            $m$ & 1 \\
            $n$ & 1 \\
                \hline
            \end{tabular}
            \caption[Global system parameters]{\textbf{Global system parameters.} The values for parameters defined in eqs.~(\ref{eq:FP_a},\ref{eq:FP_b},\ref{eq:DDE_a},\ref{eq:DDE_b}) in the main text that are maintained consistently across all simulations and numerical analysis in this work.}
            \label{tab:GlobalParameters}
    \end{table}
    
    \begin{table}[t!]
        \centering
            \begin{tabular}{ x{1.5cm} x{2cm} x{3cm} } 
				\hline
				\textbf{Figure} & \textbf{Parameter} & \textbf{Value} \\
				\hline
			  1 \& 3b,c & $D_A$ & 2.146 \textmu m${}^2$ s${}^{-1}$ \\
			   & $D_I$ & 0.386 \textmu m${}^2$ s${}^{-1}$ \\
			   & $\nu_A$ & 6.822 mol \textmu m${}^{-2}$ s${}^{-1}$ \\
			   & $\nu_I$ & 0.182 mol \textmu m${}^{-2}$ s${}^{-1}$ \\
			   & $k_A$ & 0.453 s${}^{-1}$ \\
			   & $k_I$ & 1.187 s${}^{-1}$ \\
                \hline
            1a & $L=L_\text{min}$ & 5.0 \textmu m \\
                \hline
            1b & $L$ & 27.5 \textmu m \\
                \hline
            1c & $L=L_\text{crit}$ & 47.8 \textmu m \\
                \hline
            4 & $D_A$ & 0.384 \textmu m${}^2$ s${}^-1$ \\
			   & $D_I$ & 0.00118 \textmu m${}^2$ s${}^-1$ \\
			   & $\nu_A$ & 0.371 mol \textmu m${}^{-2}$ s${}^{-1}$ \\
			   & $\nu_I$ & 0.256 mol \textmu m${}^{-2}$ s${}^{-1}$ \\
			   & $k_A$ & 0.0216 s${}^{-1}$ \\
			   & $k_I$ & 0.00531 s${}^{-1}$ \\
                \hline
            4c & $L$ & 48.7 \textmu m \\
                \hline
			  5 & $b$ & 42.990 \\
			   & $g$ & 0.0813 \\
			   & $\sigma$ & 0.175 \\
                \hline
			  5a,b & $a$ & 0.0813 \\
             & $\tau_A = \tau_I$ & 0.794 \\
                \hline
			  5c,d & $a$ & 0.0813 \\
             & $\tau_A = \tau_I$ & 79.433 \\
                \hline
            5e & $a_0$ & 0.0813 \\
                \hline
            \end{tabular}
            \caption[System parameters for figures]{\textbf{System parameters for figures.} The values for parameters defined in eqs.~(\ref{eq:FP_a},\ref{eq:FP_b},\ref{eq:DDE_a},\ref{eq:DDE_b}) in the main text that were used to generate Figs.~(\ref{fig:Oscillations},\ref{fig:NumericalHopf}b,c,\ref{fig:Hysteresis},\ref{fig:DDE}) alongside the global parameter values presented in table \ref{tab:GlobalParameters}.}
            \label{tab:IndividualParameters}
    \end{table}

    \clearpage

    \section{Calculating eigenvalues for a spatially homogeneous system} \label{sec:SI1}

    \subsection{Systems with Zero Diffusivity}

    In the absence of diffusion ($D_{A,I} = 0$), the eigenvalues of the Jacobian defined in eq.~(\ref{eq:Jacobian}) can be calculated from the trace and determinant of the Jacobian, which are equal to,
    \begin{equation}
        \text{Tr}(\underline{\underline{J}}) = -\left( \frac{k_A \, \alpha^2}{\left(\alpha+\bar{A}\right)^2} \right) -\left( \frac{k_I \, \gamma}{\gamma+\bar{A}} \right) < 0,
        \label{eq:Trace}
    \end{equation}
    \begin{equation}
        \text{Det}(\underline{\underline{J}}) = \left( \frac{k_I \, \gamma}{\gamma+\bar{A}} \right) \left[ \left( \frac{k_A \, \alpha^2}{\left(\alpha+\bar{A}\right)^2} \right) + \left( \frac{\nu_A \, h \, \beta^h \, \bar{I}^h}{\left( \gamma+\bar{A} \right) \left(\beta^h + \bar{I}^h\right)^2} \right) \right] > 0,
        \label{eq:Determinant}
    \end{equation}
    \noindent where all variable notation is the same as in the main text, and the inequalities arise from all steady-state concentrations, rates constants and Hill coefficients and constants being non-negative by definition. The eigenvalues of a matrix $\underline{\underline{M}}$ can be calculated as,
    \begin{equation}
        \lambda_\pm = \frac{\text{Tr}(\underline{\underline{M}}) \pm \sqrt{\text{Tr}(\underline{\underline{M}})^2 - 4 \, \text{Det}(\underline{\underline{M}})}}{2}. \label{eq:EigenValueDefinition}
    \end{equation}
    \noindent Substituting eqs.~(\ref{eq:Trace},\ref{eq:Determinant}) into eq.~(\ref{eq:EigenValueDefinition}) always generates eigenvalues with negative real parts, corresponding to system stability to perturbations in activator and inhibitor concentrations.

    \subsection{Systems with non-zero diffusivity}

    In the presence of diffusion ($D_{A,I} > 0$), we consider solutions of the form $\underline{\delta u} = \sum_i \underline{v}_i \exp(\lambda_i \, t + i \, q_i \, x)$, and substitute this into the equation $\partial_t \, \underline{\delta u} \simeq \underline{\underline{J}} \, \underline{\delta u}$, where $\underline{\underline{J}}$ is the Jacobian defined in eq.~(\ref{eq:Jacobian}). We define $\Lambda$ as the eigenvalue with the largest positive real part, and the corresponding wavenumber $Q$. For this solution mode the Jacobian is,
    \begin{equation}
        \underline{\underline{J}} = \begin{bmatrix}
            -\left( \frac{k_A \, \alpha^2}{\left(\alpha+\bar{A}\right)^2} \right) - D_A \, Q^2 & - \left( \frac{\nu_A \, h \, \beta^h \, \bar{I}^{h-1}}{\left(\beta^h + \bar{I}^h\right)^2} \right) \\
            \frac{k_I \, \gamma \, \bar{I}}{\left(\gamma+\bar{A}\right)^2} & -\left( \frac{k_I \, \gamma}{\gamma+\bar{A}} \right) - D_I \, Q^2
        \end{bmatrix}.
        \label{eq:JacobianWithWavenumber}
    \end{equation}
    
    \noindent Hence, $\text{Tr}(\underline{\underline{J}})<0$, and therefore $\text{Det}(\underline{\underline{J}})<0$ is necessary for $\Lambda$ to exhibit a positive real part.
    
    The determinant of eq.~(\ref{eq:JacobianWithWavenumber}) is equal to,
    \begin{equation}
        \text{Det}(\underline{\underline{J}}) = D_A \, D_I \, Q^4 + \left[ D_A \, \Gamma + D_I \, \Phi \right] Q^2 + \Gamma \left[ \Phi + \left( \frac{\nu_A \, h \, \beta^h \, \bar{I}^h}{\left( \gamma+\bar{A} \right) \left(\beta^h + \bar{I}^h\right)^2} \right) \right], \label{eq:DeterminantWithWavenumber}
    \end{equation}
    \noindent where $\Gamma = k_I \, \gamma / (\gamma+\bar{A})$, $\Phi = k_A \, \alpha^2 / (\alpha+\bar{A})^2$, and the final term is equal to the determinant defined in eq.~(\ref{eq:Determinant}), which we denote $J_0$. We next calculate the value of $Q$ that minimises eq.~(\ref{eq:DeterminantWithWavenumber}) and satisfies $d\text{Det}(\underline{\underline{J}})/dQ = 0$, corresponding to the eigenvalue with the most positive real part according to eq.~(\ref{eq:EigenValueDefinition}). The minimum value of $Q$ satisfies the equation,
    \begin{equation}
        Q_\text{min}^2 = -\left( \frac{D_A \, \Gamma + D_I \, \Phi}{2 \, D_A \, D_I} \right),
    \end{equation}
    \noindent but is imaginary and non-physical. We extract the wavenumbers $Q_0$ corresponding to where $\text{Det}(\underline{\underline{J}})=0$ to test whether any real values of the wavenumber can generate negative values of the determinant. We find that,
    \begin{equation}
        Q_0^2 = Q_\text{min}^2 \pm \sqrt{Q_\text{min}^4 - \left( \frac{J_0}{D_A \, D_I} \right)} < 0,
        \label{eq:Q0}
    \end{equation}
    \noindent where the inequality is the result of $Q_\text{min}^2 < 0$ and $J_0 / D_A \, D_I > 0$. This result means that there are no real values of the wavenumber $Q$ that generate a negative determinant, such that the system must be stable for all system parameters.

    \section{Analysing the simple DDE model} \label{sec:SI2}

    We test the stability of the non-dimensionalised DDEs defined in eqs.~(\ref{eq:DDE_a},\ref{eq:DDE_b}) using the framework by \cite{glass2021nonlinear} that follows similar methods as used in Appendix \ref{sec:SI1}. We assume small perturbations of the forms $\underline{\delta u} = (\delta X, \delta Y) = \sum_i \underline{v}_i \exp(\lambda_i \, \tau)$ around the steady-state $\bar{u} = (\bar{X}, \bar{Y})$. These steady-states are the solutions of the implicit equations,
    \begin{align}
        \bar{X} &= g \left( Y - 1 \right)^{1/n}, \label{eq:SSX} \\
        \bar{Y} &= b \left[ \left( \frac{a^m + \bar{X}^m}{a^m \, \bar{X}} \right) - 1 \right] ^ {1/h}, \label{eq:SSY}
    \end{align}
    \noindent where all variable notation is the same as in the main text. We compare these steady-states to the numerical solutions of eqs.~(\ref{eq:DDE_a},\ref{eq:DDE_b}) in the left column of Fig.~(\ref{fig:DDE}a) (see Appendix \ref{sec:SIMethods} for methods).
    
    The linearised forms of eqs.~(\ref{eq:DDE_a},\ref{eq:DDE_b}) can be written,
    \begin{equation}
        \dot{\delta X} (\tau) = \sigma \, a^m \left( \frac{(m-1) \bar{X}^m - a^m}{\left( a^m + \bar{X}^m \right)^2} \right) \delta X(\tau) - \sigma\left( \frac{h \, b^h \, \bar{Y}^{h-1}}{\left( b^h + \bar{Y}^h \right)^2} \right) \delta Y(\tau-\tau_I),
    \label{eq:DDELin_a}
    \end{equation}
    \begin{equation}
        \dot{\delta Y} (\tau) = \left( \frac{n \, g^n \, \bar{X}^{n-1} \, \bar{Y}}{\left( g^n + \bar{X}^n \right)^2} \right) \delta X(\tau-\tau_A) - \left( \frac{g^n}{g^n + \bar{X}^n} \right) \delta Y(\tau), 
    \label{eq:DDELin_b}
    \end{equation}
    \noindent which have a similar form to $\partial_\tau \, \underline{\delta u} \simeq \underline{\underline{J}} \, \underline{\delta u}$ except for the explicit inclusion of delays. For this reason, eqs.~(\ref{eq:DDELin_a},\ref{eq:DDELin_b}) cannot be written in matrix form, and we instead define the `effective' Jacobian,
    \begin{equation}
        \underline{\underline{H}} = \begin{bmatrix}
            \sigma \, a^m \left( \frac{(m-1) \bar{X}^m - a^m}{\left( a^m + \bar{X}^m \right)^2} \right) & - \sigma \left( \frac{h \, b^h \, \bar{Y}^{h-1}}{\left( b^h + \bar{Y}^h \right)^2} \right) \\
            \left( \frac{n \, g^n \, \bar{X}^{n-1}}{\left( g^n + \bar{X}^n \right)^2} \right) & - \left( \frac{g^n}{g^n + \bar{X}^n} \right)
        \end{bmatrix}.
        \label{eq:DDEJacobian}
    \end{equation}
    \noindent We will use the notation that $H_{i,j}$ are the components of this effective Jacobian throughout the remainder of our analysis.
    
    We are interested in the value of $\lambda_i$ that first exhibits a real part greater than zero, denoted $\Lambda$, and its corresponding amplitude vector $\underline{V}$. Substituting the solution mode $\underline{\delta u} = (\delta X, \delta Y) = \underline{V} \exp(\Lambda \, \tau)$ into eqs.~(\ref{eq:DDELin_a},\ref{eq:DDELin_b}) generates the pair of characteristic equations defined in eqs.~(\ref{eq:Characteristic_a},\ref{eq:Characteristic_b}) in the main text. Re-arranging these equations and multiplying them together eliminates the amplitudes $V_{1,2}$, resulting in the equation,
    \begin{equation}
        \Lambda^2 - \left( H_{1,1} + H_{2,2} \right) \Lambda + H_{1,1} \, H_{2,2} =  H_{1,2} \, H_{2,1} \, \exp\left( - \Lambda \, \left( \tau_A + \tau_I \right) \right).
        \label{eq:SingleCharacteristic}
    \end{equation}
    \noindent We can now investigate the onset of non-oscillatory (saddle-node, pitchfork etc.) or Hopf bifurcations by substituting $\Lambda = 0$ or $\Lambda = i \, \omega$, respectively, into eq.~(\ref{eq:SingleCharacteristic}).

    \subsection{Requirements for a non-oscillatory bifurcation}

    Substituting $\Lambda = 0$ into eq.~(\ref{eq:SingleCharacteristic}) generates the relation that non-oscillatory bifurcations arise when $H_{1,1} \, H_{2,2} = H_{1,2} \, H_{2,1}$. In eq.~(\ref{eq:DDEJacobian}), the signs of each of these elements is largely controlled by the values of the constants $m$, $n$ and $h$, with the exception that $H_{2,2} < 0$ for all system parameters. The signs of the remaining elements are: $\text{sgn}(H_{1,1}) = \text{sgn}((m-1) \bar{X}^m - a^m)$; $\text{sgn}(H_{1,2}) = \text{sgn}(-h)$; and $\text{sgn}(H_{2,1}) = \text{sgn}(n)$. In the case of $m=n=1$, corresponding to the feedback of the full model defined in eqs.~(\ref{eq:FP_a},\ref{eq:FP_b}) in the main text, then $\text{sgn}(H_{1,1}) < 0$ and $\text{sgn}(H_{2,1}) > 0$, such that $\text{sgn}(H_{1,2}) = \text{sgn}(-h) > 0$ is necessary for a non-oscillatory bifurcation. This is not true for the simulations of the full system shown in Figs.~(\ref{fig:Oscillations}-\ref{fig:Hysteresis}), which use the value of $h=4$.

    \subsection{Requirements for a Hopf bifurcation}

    We first define the function,
    \begin{equation}
        f(H_{i,j}) = \begin{cases} 
            0 & \text{if } \text{sgn}\left( H_{i,j} \right) > 0, \\
            \pi & \text{if } \text{sgn}\left( H_{i,j} \right) < 0,
        \end{cases}
        \label{eq:ExtractImaginaryFunction}
    \end{equation}
    \noindent which converts the sign of $H_{i,j}$ into a value that returns the sign again when substituted into the equation $\exp(i \, f(H_{i,j}))$. As in the previous section, the signs of each element of the effective Jacobian can be written: $\text{sgn}(H_{1,1}) = \text{sgn}((m-1) \bar{X}^m - a^m)$; $\text{sgn}(H_{1,2}) = \text{sgn}(-h)$; $\text{sgn}(H_{2,1}) = \text{sgn}(n)$; and $\text{sgn}(H_{2,2}) < 0$. Substituting $\Lambda = i \, \omega$ into eq.~(\ref{eq:SingleCharacteristic}) and using eq.~(\ref{eq:ExtractImaginaryFunction}) generates the equations,
    \begin{equation}
    \begin{aligned}
        & - \omega^2 - \omega | H_{1,1} | \cos\left( f(H_{1,1}) + \left( \frac{(4j + 1)\pi}{2} \right) \right) + | H_{1,1} | | H_{2,2} | \cos\left( f(H_{1,1}) + \pi \right)  \\
        & \qquad\qquad\qquad\qquad\qquad\qquad\qquad\qquad\qquad = | H_{1,2} | | H_{2,1} | \cos\left( f(H_{1,2}) + f(H_{2,1}) - \omega \left( \tau_A + \tau_I \right) \right),
    \end{aligned}
    \label{eq:RealPart}
    \end{equation}
    \begin{equation}
    \begin{aligned}
        & - \omega | H_{1,1} | \sin\left( f(H_{1,1}) + \left( \frac{(4j + 1)\pi}{2} \right) \right) + \omega | H_{2,2} | + | H_{1,1} | | H_{2,2} | \sin\left( f(H_{1,1}) + \pi \right)  \\
        & \qquad\qquad\qquad\qquad\qquad\qquad\qquad\qquad\qquad = | H_{1,2} | | H_{2,1} | \sin\left( f(H_{1,2}) + f(H_{2,1}) - \omega \left( \tau_A + \tau_I \right) \right),
    \end{aligned}
    \label{eq:ImaginaryPart}
    \end{equation}
    \noindent where eq.~(\ref{eq:RealPart}) corresponds to the real parts of eq.~(\ref{eq:SingleCharacteristic}) and eq.~(\ref{eq:ImaginaryPart}) corresponds to the imaginary parts, and we have used that $i = \exp(i ((4j + 1)\pi / 2))$ for any integer $j$.

    In the case of $m=1$, $n=1$ and $h=4$, corresponding to the simulations of the full system shown in Figs.~(\ref{fig:Oscillations}-\ref{fig:Hysteresis}), eqs.~(\ref{eq:RealPart},\ref{eq:ImaginaryPart}) become,
    \begin{equation}
        - \omega^2 + | H_{1,1} | | H_{2,2} | = - | H_{1,2} | | H_{2,1} | \cos\left( \omega \left( \tau_A + \tau_I \right) \right),
        \label{eq:RealPartSimplified}
    \end{equation}
    \begin{equation}
        \omega \left( | H_{1,1} | + | H_{2,2} | \right) =  | H_{1,2} | | H_{2,1} | \sin\left( \omega \left( \tau_A + \tau_I \right) \right).
        \label{eq:ImaginaryPartSimplified}
    \end{equation}
    \noindent Dividing eq.~(\ref{eq:ImaginaryPartSimplified}) by eq.~(\ref{eq:RealPartSimplified}) and re-arranging the result generates the equation for the combined delay $\tau_A + \tau_I$ defined in eq.~(\ref{eq:AnalyticalThreshold1}) in the main text. Substituting eq.~(\ref{eq:AnalyticalThreshold1}) into eq.~(\ref{eq:RealPartSimplified}) and utilising the relation $\cos \left[ \tan^{-1} \left( Z \right) \right] = 1 / \sqrt{1 + Z^2}$ generates the equation for the oscillation frequency at the critical point $\omega$ defined in eq.~(\ref{eq:AnalyticalThreshold2}) in the main text. Since $\tau_A + \tau_I > 0$ by definition, the argument of the arctan function in eq.~(\ref{eq:AnalyticalThreshold1}) must also be greater than zero, or equivalently $\omega^2 > |H_{1,1}| |H_{2,2}|$. Similarly, in order for $\omega$ to be real, we find that $|H_{1,2}| |H_{2,1}| > |H_{1,1}| |H_{2,2}|$.

    \section*{Data, Materials, and Software Availability}
    
    The Julia code used to generate all computational results in this work can be found at: \url{https://www.github.com/LMosby/SpatiotemporalOscillations}.

    \section*{Author Contributions}

    \textbf{L.S. Mosby}: designed research; performed research; contributed analytical and numerical tools; analysed data; wrote paper. \textbf{M.P. Dalwadi}: performed research; contributed analytical and numerical tools; reviewed and edited paper. \textbf{Z. Hadjivasiliou}: designed research; performed research; contributed analytical and numerical tools; reviewed and edited paper.
    
    \section*{Acknowledgements}
    
    This work was supported by the Francis Crick Institute, which receives its core funding from Cancer Research UK, the UK Medical Research Council, and Wellcome Trust. For the purpose of open access, the author has applied a CC BY public copyright licence to any author accepted manuscript arising from this submission.
    
    \end{spacing}

    \clearpage

    \bibliographystyle{unsrt}
    \bibliography{bib}

@article{fussmann2000crossing,
  title={Crossing the {Hopf} bifurcation in a live predator-prey system},
  author={Fussmann, Gregor F and Ellner, Stephen P and Shertzer, Kyle W and Hairston Jr, Nelson G},
  journal={Science},
  volume={290},
  number={5495},
  pages={1358--1360},
  year={2000},
  publisher={American Association for the Advancement of Science}
}

@article{martinez2018bistable,
  title={Bistable emergence of oscillations in growing Bacillus subtilis biofilms},
  author={Martinez-Corral, Rosa and Liu, Jintao and S{\"u}el, G{\"u}rol M and Garcia-Ojalvo, Jordi},
  journal={Proc. Natl. Acad. Sci. U.S.A.},
  volume={115},
  number={36},
  pages={E8333--E8340},
  year={2018},
  publisher={National Academy of Sciences}
}

@article{di2018landau,
  title={Landau--Ginzburg theory of cortex dynamics: Scale-free avalanches emerge at the edge of synchronization},
  author={Di Santo, Serena and Villegas, Pablo and Burioni, Raffaella and Mu{\~n}oz, Miguel A},
  journal={Proc. Natl. Acad. Sci. U.S.A.},
  volume={115},
  number={7},
  pages={E1356--E1365},
  year={2018},
  publisher={National Academy of Sciences}
}

@article{biga2021dynamic,
  title={A dynamic, spatially periodic, micro-pattern of HES5 underlies neurogenesis in the mouse spinal cord},
  author={Biga, Veronica and Hawley, Joshua and Soto, Ximena and Johns, Emma and Han, Daniel and Bennett, Hayley and Adamson, Antony D and Kursawe, Jochen and Glendinning, Paul and Manning, Cerys S and others},
  journal={Mol. Syst. Biol.},
  volume={17},
  number={5},
  pages={MSB20209902},
  year={2021},
  publisher={Springer}
}

@article{kuniya2025hopf,
  title={Hopf bifurcation in a time-delayed multi-group {SIR} epidemic model for population behavior change},
  author={Kuniya, Toshikazu},
  journal={J. Math. Anal. Appl.},
  pages={130061},
  year={2025},
  publisher={Elsevier}
}

@article{glass2021nonlinear,
  title={Nonlinear delay differential equations and their application to modeling biological network motifs},
  author={Glass, David S and Jin, Xiaofan and Riedel-Kruse, Ingmar H},
  journal={Nat. Commun.},
  volume={12},
  number={1},
  pages={1788},
  year={2021},
  publisher={Nature Publishing Group UK London}
}

@article{krause2018heterogeneity,
  title={Heterogeneity induces spatiotemporal oscillations in reaction-diffusion systems},
  author={Krause, Andrew L and Klika, V{\'a}clav and Woolley, Thomas E and Gaffney, Eamonn A},
  journal={Phys. Rev. E},
  volume={97},
  number={5},
  pages={052206},
  year={2018},
  publisher={APS}
}

@article{ridgway2023motility,
  title={Motility-induced phase separation mediated by bacterial quorum sensing},
  author={Ridgway, Wesley JM and Dalwadi, Mohit P and Pearce, Philip and Chapman, S Jonathan},
  journal={Phys Rev Lett},
  volume={131},
  number={22},
  pages={228302},
  year={2023},
  publisher={APS}
}

@article{dalwadi2023universal,
  title={Universal dynamics of biological pattern formation in spatio-temporal morphogen variations},
  author={Dalwadi, Mohit P and Pearce, Philip},
  journal={Proc R Soc A},
  volume={479},
  number={2271},
  pages={20220829},
  year={2023},
  publisher={The Royal Society}
}

@article{mosby2026position,
  title={Position-dependent feedback drives scaling and robustness of morphogen gradients},
  author={Mosby, Lewis Scott and Hadjivasiliou, Zena},
  journal={Proc. Natl. Acad. Sci. U.S.A.},
  volume={123},
  number={21},
  pages={e2521182123},
  year={2026},
  publisher={National Academy of Sciences}
}

@article{ben2010scaling,
  title={Scaling of morphogen gradients by an expansion-repression integral feedback control},
  author={Ben-Zvi, Danny and Barkai, Naama},
  journal={Proc. Natl. Acad. Sci. U.S.A.},
  volume={107},
  number={15},
  pages={6924--6929},
  year={2010},
  publisher={National Academy of Sciences}
}

@article{almuedo2018scale,
  title={Scale-invariant patterning by size-dependent inhibition of Nodal signalling},
  author={Almuedo-Castillo, Mar{\'\i}a and Bl{\"a}{\ss}le, Alexander and M{\"o}rsdorf, David and Marcon, Luciano and Soh, Gary H and Rogers, Katherine W and Schier, Alexander F and M{\"u}ller, Patrick},
  journal={Nat. Cell Biol.},
  volume={20},
  number={9},
  pages={1032--1042},
  year={2018},
  publisher={Nature Publishing Group UK London}
}

@article{martin2009cell,
  title={Cell competition, growth and size control in the \textit{Drosophila} wing imaginal disc},
  author={Mart{\'\i}n, Francisco A and Herrera, Salvador C and Morata, Gin{\'e}s},
  journal={Development},
  volume={136},
  number={22},
  pages={3747--3756},
  year={2009},
  publisher={The Company of Biologists Limited}
}

@article{zhang2022fgf8,
  title={Fgf8 dynamics and critical slowing down may account for the temperature independence of somitogenesis},
  author={Zhang, Weiting and Scerbo, Pierluigi and Delagrange, Marine and Candat, Virginie and Mayr, Vanessa and Vriz, Sophie and Distel, Martin and Ducos, Bertrand and Bensimon, David},
  journal={Commun. Biol.},
  volume={5},
  number={1},
  pages={113},
  year={2022},
  publisher={Nature Publishing Group UK London}
}

@article{aguilar2018critical,
  title={Critical point in self-organized tissue growth},
  author={Aguilar-Hidalgo, Daniel and Werner, Steffen and Wartlick, Ortrud and Gonz{\'a}lez-Gait{\'a}n, Marcos and Friedrich, Benjamin M and J{\"u}licher, Frank},
  journal={Phys. Rev. Lett.},
  volume={120},
  number={19},
  pages={198102},
  year={2018},
  publisher={APS}
}

@article{autorino2025closed,
  title={A closed feedback between tissue phase transitions and morphogen gradients drives patterning dynamics},
  author={Autorino, Camilla and Khoromskaia, Diana and Harari, Louise and Floris, Elisa and Booth, Harry and Pallares-Cartes, Cristina and Petrasiunaite, Vesta and Dorrity, Michael and Corominas-Murtra, Bernat and Hadjivasiliou, Zena and Petridou, Nicoletta I.},
  journal={bioRxiv},
  pages={2025--06},
  year={2025},
  publisher={Cold Spring Harbor Laboratory}
}

@article{turing1952chemical,
  title={The Chemical Basis of Morphogenesis},
  author={Turing, Alan Mathison},
  journal={Philos. Trans. R. Soc. B},
  volume={237},
  number={641},
  pages={37--72},
  year={1952},
  publisher={The Royal Society}
}

@article{alonso2025amplification,
  title={Amplification through local critical behavior in the mammalian cochlea},
  author={Alonso, Rodrigo G and Gianoli, Francesco and Fabella, Brian and Hudspeth, AJ},
  journal={Proc. Natl. Acad. Sci. U.S.A.},
  volume={122},
  number={29},
  pages={e2503389122},
  year={2025},
  publisher={National Academy of Sciences}
}

@article{macnamara2016diffusion,
  title={Diffusion driven oscillations in gene regulatory networks},
  author={Macnamara, Cicely K and Chaplain, Mark AJ},
  journal={J. Theor. Biol.},
  volume={407},
  pages={51--70},
  year={2016},
  publisher={Elsevier}
}

@article{simons2026cell,
  title={Cell cycle criticality as a mechanism for robust cell population control},
  author={Simons, Benjamin D and Karin, Omer},
  journal={Mol. Syst. Biol.},
  volume={22},
  number={2},
  pages={241--258},
  year={2026},
  publisher={Springer}
}

@article{raspopovic2014digit,
  title={Digit patterning is controlled by a Bmp-Sox9-Wnt Turing network modulated by morphogen gradients},
  author={Raspopovic, Jelena and Marcon, Luciano and Russo, Laura and Sharpe, James},
  journal={Science},
  volume={345},
  number={6196},
  pages={566--570},
  year={2014},
  publisher={American Association for the Advancement of Science}
}

@article{bastida2009bmp,
  title={A {BMP-Shh} negative-feedback loop restricts Shh expression during limb development},
  author={Bastida, Ma F{\'e}lix and Sheth, Rushikesh and Ros, Maria A},
  journal={Development},
  volume={136},
  number={22},
  pages={3779--3789},
  year={2009},
  publisher={The Company of Biologists Limited}
}

\end{document}